\documentclass[journal]{new-aiaa}

\usepackage{bm}
\usepackage{amsmath}
\usepackage{subfigure}
\usepackage{comment}
\usepackage{footnpag}			      	% make footnote symbols restart on each page
\usepackage{subfigure}   
\usepackage{multicol}
\usepackage{multirow}
\usepackage{booktabs}
\usepackage{pstool}             % for changing text in eps graphics

\usepackage{longtable}
\newcommand{\mbf}{\mathbf}

\begin{document}

\title{Threat Level Estimation From Possible Break-Up Events In LEO\footnote{Paper presented at the 2024 AIAA SciTech Forum and Exposition (8-12 January 2024, Orlando, FL, USA). Paper number: AIAA-2024-1065.}}

\author{Simone Servadio\footnote{Assistant Professor, Department of Aerospace Engineering, Iowa State University, Ames, IA, 50011, USA. email: servadio@iastate.edu}}
\affil{Iowa State University, Ames IA, 50011, USA}
\author{Daniel Jang\footnote{PhD Candidate, Department of Aeronautics and Astronautics, Massachusetts Institute of Technology, MA 02139, USA. email: djang@mit.edu}}
\author{ Richard Linares\footnote{Rockwell International Career Development Professor, Associate Professor of Aeronautics and Astronautics, Department of Aeronautics and Astronautics, Massachusetts Institute of Technology, Cambridge, MA 02139, USA. email: linaresr@mit.edu}}
\affil{Massachusetts Institute of Technology, Cambridge, MA, 02139, USA}

 \maketitle
 
\begin{abstract}
The NASA Standard Break-Up Model models collisions and explosions in space, which identifies the future distribution of debris. Given a possible break-up event, this work analyses the threat posed by the generated debris in the orbit of an asset spacecraft. Using the Koopman Operator solution of the $J_2$ perturbed two-body dynamics, the family of all possible orbits that cross the asset's pathway is identified and parameterized according to their velocity. The threat level assessment of a collision onto the asset is estimated considering the intersection between the velocities of the break-up model and the Koopman transfer solutions. 
% 99/100 words
\end{abstract}

%%%%%%%%%%%%%%%%%%%%%%%%%%%%%%%%%%%%%%%%%%%%%%%%%%%%%%%%%%%%%%%%%%%%%%%%%%%%%%%
%%%%%%%%%%%%%%%%%%%%%%%%%%%%%%%%%%%%%%%%%%%%%%%%%%%%%%%%%%%%%%%%%%%%%%%%%%%%%%%
%%%%%%%%%%%%%%%%%%%%%%%%%%%%%%%%%%%%%%%%%%%%%%%%%%%%%%%%%%%%%%%%%%%%%%%%%%%%%%
\section*{Nomenclature}
{\renewcommand\arraystretch{1.0}
\noindent\begin{longtable*}{@{}l @{\quad=\quad} l@{}}
$a$  & Semimajor Axis  \\
$d_{tr}$  & Tractable Diameter \\
$e$  & Eccentricity  \\
$E$  & Specific Energy of the Orbit  \\
$\mbf g()$  & Observable Function  \\
$h$  & Altitude  \\
$H$  & Tracking Relative Altitude   \\
$\mbf H$  & Koopman Observable Matrix   \\
$i$  & Inclination  \\
$J_2$  & Earth Oblateness Perturbation Constant   \\
$\mbf K$  & Koopman  Matrix   \\
$\hat l$  & Normalized Length  \\
$\mbf L()$  & Legendre Polynomials  \\
$\hat m$  & Normalized Mass \\
$m_p$  & Projectile Mass \\
$m_t$  & Target Mass \\
$\mathcal M_{\theta_0 \rightarrow \theta_f}$  & State Transition Polynomial Map  \\
$\mathcal M^{DA}_{t_0 \rightarrow t_f}$  & Differential Algebra State Transition Polynomial Map  \\
$\mathcal M_{\theta_0 \rightarrow \theta_f}^{KO}$  & Koopman Operator State Transition Polynomial Map  \\
$n_f$  & Number of Fragments  \\
$r$ & Radial Distance \\
$\mbf r$ & Position Vector \\
$p_{r}$ & Conjugate Momentum - Radial Distance\\
$p_{\phi}$ & Conjugate Momentum - Latitude \\
$p_{\lambda}$ & Conjugate Momentum - Longitude\\
$R_{\oplus}$ & Earth's Radius \\
$\mathcal T()$  & Cartesian/Polar Transformation  \\
$\mbf v$  & Velocity Vector  \\
$V$  & Volume  \\
$\mbf V$  & Eigenvector Matrix  \\
$\mathcal W^{DA}_{t_f \rightarrow t_0}$  & Inverted Differential Algebra State Transition Polynomial Map  \\
$\mbf x$  & State of the System \\
$\varphi$ & Latitude \\
$\lambda$ & Longitude \\
$\Lambda $  & Eigenvalues Matrix  \\
$\mu$ & Earth's gravitational constant \\
$\mu_\nu$ & NASA SMB velocities PDF mean \\
$\Xi$ & Transfer Orbits Surface \\
$\sigma_\nu$ & NASA SMB velocities PDF standard deviation  \\
$\theta$ & True Anomaly \\
$\tau$ & Convex Hull Threshold  \\
$\omega$ & Argument of the Periapsis \\
$\Omega$ & Right Ascension of the Ascending Node \\

 % \multicolumn{2}{@{}l}{Subscripts}\\
 % $G$ & generator body\\
\end{longtable*}}

%%%%%%%%%%%%%%%%%%%%%%%%%%%%%%%%%%%%%%%%%%%%%%%%%%%%%%%%%%%%%%%%%%%%%%%%%%%%%%%
%%%%%%%%%%%%%%%%%%%%%%%%%%%%%%%%%%%%%%%%%%%%%%%%%%%%%%%%%%%%%%%%%%%%%%%%%%%%%%%
%%%%%%%%%%%%%%%%%%%%%%%%%%%%%%%%%%%%%%%%%%%%%%%%%%%%%%%%%%%%%%%%%%%%%%%%%%%%%%%
\section{Introduction}
Nowadays, more than ever, the space race is reaching its peak. In the last few years, humanity has passed from launching an average of a single satellite each month to launching large constellations in one year. The Low Earth Orbit (LEO) space is becoming crowded, and safety has become the number one priority to ensure the sustainability of the space environment \cite{d2023novel}. Regardless of the technological advantages, failures in post-mission disposal or avoidance maneuvers are doomed to happen due to the stochastic nature of the systems. Therefore, it is critical for governments and private companies to become aware of the danger posed by space debris and other satellites such that they can plan collision avoidance maneuvers whenever needed. 

Past events support the need for collision awareness and its consequences. A collision in space has long-lasting effects as debris are doomed to continue to orbit around Earth for centuries. Therefore, in order to ensure a sustainable space environment with safety as the number one priority, the need for precise and accurate payload maneuvers and increased awareness of the threat posed by space debris is fundamental. Some debris orbiting in LEO are a constant threat to main constellations due to their physical characteristic and orbit parameters \cite{servadio2023risk}. In the middle of the last century, with the state of space exploration and human space activities, a population of nonfunctional objects, the so-called space debris, has started to grow exponentially. The first recorded explosion occurred in 1961 when a rocket's upper stage created 298 trackable fragments with sizes larger than 10 cm in LEO \cite{KlinkradBook2006}. Since then, the number of trackable debris in LEO has continuously increased and the issue of possible collision has become a constant threat to satellites, such that a list of the debris Two Line Elements (TLE) needs to be updated regularly to help spacecraft prepare their avoidance maneuvers. 

Several critical events have shown that awareness is the first step to ensuring safety in the space environment. For example, consider the 2009 collision between the satellite Cosmos 2251 and Iridium 33 \cite{kelso2009analysis}, which forced the whole Iridium constellation to perform evasive maneuvers on a weekly basis to avoid a ripple collision effect. Similarly, the impact of a single individual fragment on European Satellite-1A in 2016 could have degenerated into a cloud of debris if the debris' dimensions were bigger than 1 cm \cite{krag20171}.  SpaceX has reported\footnote{https://www.space.com/starlink-satellite-conjunction-increase-threatens-space-sustainability} that their Starlink constellation of 5000 satellites has had to perform more than 25,000 collision avoidance maneuvers (CAM) within a span of 6 months between Dec. 1, 2022, and May 31, 2023.  With the increased density along certain parts of LEO and the growing number of debris from explosions, collisions, and launches, the computational burden of identifying potential collisions quickly and accurately is paramount.  

This paper provides a methodology to improve the safety of a spacecraft when analyzing close conjunctions in terms of worst-case scenarios. The software algorithm proposed here analyzes the danger and threat level posed by a possible collision with a specific satellite orbiting in LEO. Given a break-up event, the new methodology analyzes the distribution of the debris created, their dispersion, and the family of orbits they populate to assess if they pose a collision threat to the asset (satellite of interest). Using the NASA Standard Break-Up model EVOLVE4 \cite{nasaSTMevolve4}, resident space objects (RSO) from collisions and explosions are modeled in terms of debris mass, size, and velocity. Using high-order state transition polynomial maps (STPM) created by the Koopman Operator (KO) and Differential Algebra (DA), the algorithm will be able to determine if any of the debris generated by the break-up event represents a threat to the asset, analyzing if any part of the generated cloud of debris crosses the orbit of the satellite to be protected. 

After selecting a possible collision between two RSOs that have a close conjunction, NASA's model generates a database of debris with a corresponding set of two line elements (TLE). Therefore, the proposed technique analyses whether there is any possibility of an intersection between the orbits of the debris and the orbit of the asset.  The methodology is tested against the historical collision between Iridium 33 and Kosmos 2251 in 2009 to a hypothetical Starlink satellite at 550 km altitude.

%%%%%%%%%%%%%%%%%%%%%%%%%%%%%%%%%%%%%%%%%%%%%%%%%%%%%%%%%%%%%%%%%%%%%%%%%%%%%%%
%%%%%%%%%%%%%%%%%%%%%%%%%%%%%%%%%%%%%%%%%%%%%%%%%%%%%%%%%%%%%%%%%%%%%%%%%%%%%%%
%%%%%%%%%%%%%%%%%%%%%%%%%%%%%%%%%%%%%%%%%%%%%%%%%%%%%%%%%%%%%%%%%%%%%%%%%%%%%%%
\section{Koopman Operator Solution}
The Koopman Operator (KO) solution of the two-body dynamics affected by gravitational $J_2$ perturbations has been selected as one of the tools to propagate the state of the system \cite{servadio2021koopman,servadio2022dynamics}. This selection comes with various advantages over classic numerical propagation. First, the KO is applied to the zonal formulation of the dynamics, where the time regularization changes the independent variable to the true anomaly spanned by the secondary body from its initial condition. Secondly, the KO generates a State Transition Polynomial Map (STPM) that directly connects an initial condition at a given anomaly to its final condition after a well-selected angle, such that the propagation is done by means of polynomial evaluation. 

%%%%%%%%%%%%%%%%%%%%%%%%%%%%%%%%%%%%%%%%%%%%%%%%%%%%%%%%%%%%%%%%%%%%%%%%%%%%%%%
\subsection{The Zonal Harmonics}
The perturbed two-body problem can be represented using zonal harmonics. The zonal harmonics problem has been solved considering a new set of orbital elements \cite{arnas2021approximate}. Considering the spherical representation of the state of the system, in terms of radius, angles, and momenta, the transformation to the new set is: 
\begin{equation}
\begin{aligned}
\Lambda &=\sqrt{\frac{R_{\oplus}}{\mu}}\left(\frac{p_\theta}{r}-\frac{\mu}{p_\theta}\right)\\
\eta &=p_r \sqrt{\frac{R_{\oplus}}{\mu}} \\
s &=\sin (\varphi) \\
\gamma &=\frac{p_{\varphi}}{p_\theta} \cos (\varphi)  \\
\kappa &=\sqrt{\frac{\mu R_{\oplus}}{p_{\varphi}^2+\frac{p_\lambda^2}{\cos ^2(\varphi)}}} \\
\beta &=\lambda-\arcsin \left(\tan (\varphi) \sqrt{\frac{p_\lambda^2}{p_\theta^2-p_\lambda^2}}\right)  \\
\chi &=\frac{p_\lambda}{p_\theta^4} \frac{\left(\mu R_{\oplus}\right)^{3 / 2} }{\sin ^2(\varphi)+\frac{p_\varphi^2}{p_\theta^2} \cos ^2(\varphi)} \\
\rho &=\frac{p_\lambda}{p_\theta}
\end{aligned}
\label{eq:orbitalelements}
\end{equation}
where $\mu$ is the gravitational constant, $R_{\oplus}$is Earth's equatorial radius, $r$ is the radial distance between the spacecraft and the center of the Earth, $\varphi$ is the latitude, and $\lambda$ is the longitude. Moreover, the conjugate momenta are expressed as 
\begin{equation}
\begin{aligned}
p_r &= \dot r\\
p_\phi &= r^2 \dot \phi\\
p_\lambda &= r^2 \dot \lambda\cos^2( \phi)\\ 
\end{aligned}
\end{equation}
Lastly, variable
\begin{equation}
p_\theta=\sqrt{p_{\varphi}^2+\frac{p_\lambda^2}{\cos ^2(\varphi)}}
\end{equation}
is the angular momentum of the orbit. After the time regularization 
\begin{equation}
\frac{d \theta}{d t}=\frac{p_\theta}{r^2}
\end{equation}
the equations of motion can be expressed as 
\begin{equation}
\begin{aligned}
\frac{d \Lambda}{d \theta} &=-\eta-3 J_2 s \gamma \kappa^3(\Lambda+\kappa)(\Lambda+2 \kappa) \\
\frac{d \eta}{d \theta} &=\Lambda+\frac{3}{2} J_2 \kappa^3(\Lambda+\kappa)^2\left(3 s^2-1\right) \\
\frac{d s}{d \theta} &=\gamma \\
\frac{d \gamma}{d \theta} &=-s-3 J_2 s \rho^2 \kappa^3(\Lambda+\kappa) \\
\frac{d \kappa}{d \theta} &=3 J_2 s \gamma \kappa^4(\Lambda+\kappa) \\
\frac{d \beta}{d \theta} &=-3 J_2 s^2 \chi(\Lambda+\kappa) \\
\frac{d \chi}{d \theta} &=12 J_2 s \gamma \chi \kappa^3(\Lambda+\kappa)+6 J_2 s \rho \chi^2(\Lambda+\kappa) \\
\frac{d \rho}{d \theta} &=3 J_2 s \gamma \rho \kappa^3(\Lambda+\kappa)
\end{aligned}
\label{eqn:elemwithj2}
\end{equation}
The linear contributions to these variables that do not have any terms proportional to $J_2$ give the system of the classical two-body problem without $J_2$ effects. The full derivation of the equations is offered in the references, and it shows how the transformation Eq. \eqref{eq:orbitalelements} has been applied to the spherical Hamiltonian, such that the equations of motion are obtained through Hamilton equations.

%%%%%%%%%%%%%%%%%%%%%%%%%%%%%%%%%%%%%%%%%%%%%%%%%%%%%%%%%%%%%%%%%%%%%%%%%%%%%%%
\subsection{The Koopman Operator State Transition Polynomial Map}
Given a specific initial condition, the KO propagates the dynamics up to a specific point. This work uses the Koopman Operator as a tool to derive the KO polynomial map based on orthogonal polynomials, and it describes a novel use of such a solution to evaluate the set of velocities solving the Two Point Boundary Value Problem (TPBVP). Therefore, the derivation and application of the operator follow the previously published methodology described in previous works, both by the authors and by other scholars, and it will not be reported in the paper \cite{servadio2022dynamics, servadio2023uncertainty, hofmann2022advances, KOinversejournal, pasiecznik2022lambert}. 

The main idea of the Koopman Operator is to project the system's dynamics onto a set of orthogonal polynomials. After describing the rate of change of those polynomials in time, it is possible to derive the time behavior of any observable as a linear combination of the selected polynomial basis. In the general case, the selected observable is the identity observable, as it is desired to obtain the state of the system. The Koopman solution of the dynamics has the form of
\begin{equation}
    \mbf x_f = \mbf H \mbf V^{-1} \exp{(\mbf \Lambda \Delta\theta)}\mbf V \mbf L(\mbf x_0) \label{KOsol}
\end{equation}
where $\mbf L(\mbf x_0)$ indicates the value of the basis function, selected Legendre polynomials, at the given initial condition; $\mbf V$ and $\mbf \Lambda$ are the system's eigenvectors and eigenvalues, respectfully; $\Delta\theta = \theta_f - \theta_0$, with $\theta_f$ the final true anomaly; and $\mbf H$ is the observable matrix, which obtains the state as a combination of the propagated polynomials. This solution comes from the evaluation of the Koopman matrix $\mbf K$ and its eigendecomposition
\begin{equation}
    \mbf V\mbf K = \mbf \Lambda \mbf V
\end{equation}
There are different methodologies to obtain this matrix, either numerically through Extended Dynamic Mode Decomposition (EDMD) \cite{williams2015data} or analytically using the Galerkin method. This paper implements the latter solution following the direction in \cite{servadio2021koopman}. By looking at Eq. \eqref{KOsol}, it can be noted how it is constructed. First, the basis functions are evaluated at the given initial condition, $\mbf L(\mbf x_0)$; then, they are transformed into eigenfunctions of the system due to the multiplication of $\mbf V$. They represent the so-called eigenfunctions of the system $\boldsymbol \phi(\theta_0) = \mbf V \mbf L(\mbf x_0)$. Since the dynamics of the system are diagonal with respect to the eigenfunctions, their propagation is computed by the multiplication of the $\exp{(\mbf \Lambda \Delta\theta)}$ component, which corresponds to the value of the eigenfunction at the final true anomaly, $\boldsymbol \phi(\theta_f)$. The eigenfunctions are transformed back to the basis functions via the inverse transformation $\mbf V^{-1}$. Lastly, the observable is extracted by multiplying the basis functions by the observable matrix $\mbf H$, which identifies the linear combination of how much of each polynomial constitutes the selected observable. The following set of equations explains, step by step, the process described so far without going too much into detail on how the KO solution is extracted:
\begin{align}
\mbf L(\theta_0) = \mbf L(\mbf x(\theta_0)) \quad &\text{Basic Functions at }\theta_0 \nonumber\\
\boldsymbol \phi(\theta_0) = \mbf V \mbf L(\mbf x(\theta_0)) \quad &\text{Eigenfunctions at }\theta_0 \nonumber\\
\boldsymbol \phi(\theta_f) = \exp{(\mbf \Lambda \Delta\theta)}\mbf V \mbf L(\mbf x(\theta_0)) \quad &\text{Eigenfunctions at }\theta_f \nonumber\\
\mbf L(\theta_f) = \mbf V^{-1} \exp{(\mbf \Lambda \Delta\theta)}\mbf V \mbf L(\mbf x(\theta_0)) \quad &\text{Basic Functions at }\theta_f \nonumber\\
\mbf g(\theta_f) = \mbf H \mbf V^{-1} \exp{(\mbf \Lambda \Delta\theta)}\mbf V \mbf L(\mbf x(\theta_0)) \quad &\text{Observable at }\theta_f \nonumber
\end{align}
Thanks to this formulation, it can be noted that by leaving the initial condition as a generic variable and merging together the contribution of each polynomial basis function, a comprehensive polynomial map is created. This map, made of polynomials, has the initial condition of the state as its variable; the coefficients of each monomial have been calculated directly from the KO map, and it outputs the desired observable whenever evaluated at a given state. This type of dynamical system solution takes the name of State Transition Polynomial Map (STPM), and since, in this case, the coefficients of the polynomials are obtained using the Koopman Operator, we call it the Koopman Operator STPM (KOSTPM).  

The KO solution is a map that connects the initial, $\mbf x_{0,Z}$ to the final condition, $\mbf x_{f,Z}$, where the subscript $Z$ indicates the state of the system expressed according to the zonal set of variables. The KOSTPM is expressed in a compact form in the following way:
\begin{equation}
    \mbf x_{f,Z} = \mathcal M_{\theta_0 \rightarrow \theta_f}  (\mbf x_{0,Z})
\end{equation}
where $\mathcal M_{\theta_0 \rightarrow \theta_f}$ is the KOSTPM from initial $\theta_0$ to the final $\theta_f$. This map can be adjusted to work with the Cartesian representation of the state, so it is intuitive to study the effects of the change in velocity. Indeed, by calling with $\mathcal T()$ the nonlinear transformation that changes the representation of the state from zonal to Cartesian, the STPM can be modified to work with the new set of variables. 
\begin{equation}
    \mbf x_{f,C} = \mathcal T^{-1}(\mathcal M_{\theta_0 \rightarrow \theta_f}  (\mathcal T(\mbf x_{0,C}))) = \tilde{ \mathcal M}_{\theta_0 \rightarrow \theta_f} (\mbf x_{0,C}) \label{KOmap}
\end{equation} 
where $\mathcal T^{-1}()$ is the inverted transformation and $\tilde{ \mathcal M}_{\theta_0 \rightarrow \theta_f}$ is the new KOSTPM that includes the transformations. This transformation \cite{arnas2021approximate} is obtained following Eq. \eqref{eq:orbitalelements} and merging the spherical to Cartesian change of variable. The map representation of the dynamics is exploited to evaluate all the possible velocities solutions that connect two points in space, similar to an undefined Lambert's problem without the time of flight constraint.

%%%%%%%%%%%%%%%%%%%%%%%%%%%%%%%%%%%%%%%%%%%%%%%%%%%%%%%%%%%%%%%%%%%%%%%%%%%%%%%
\subsection{The KOSTPM Velocity Hyperbola}
The polynomial map defined so far is a 1-on-1 function that connects the initial to the final position. However, the initial position vector, assumed to be the point where the collision occurs, is assumed to be known and given. Similarly, the positions of the asset spacecraft that is desired to protect are also known since they represent the asset's parking orbit. It is possible to include this information in the KOSTPM to simplify the map and its dependency. That is, the map is evaluated at the given initial position, the numerical value of $\mbf{\bar r}_0$, where $\mbf{\bar x}_{0,C} = [\mbf{\bar r}_0^T \ \mbf{\bar v}_0^T]^T$. 
\begin{equation}
 \mbf x_{f,C} = \mathcal M^{KO}_{\theta_0 \rightarrow \theta_f}  (\mbf{\bar r}_0, \mbf v_0) = 
 \mathcal M^{KO}_{\mbf{\bar r}_0, \theta_0 \rightarrow \theta_f}  (\mbf v_0) \label{KOmapc}
\end{equation}
where the superscript indicates that the map has been derived using the KO approach, and the subscript indicates that it has been evaluated at the given initial position. However, Eq. \eqref{KOmapc} propagates the dynamics to the full Cartesian state, including the velocity values at the final time. Only the first half of the full map is needed, the part that connects the initial velocity to the final position. 
\begin{equation}
    \mbf x_{f,C} = 
    \begin{bmatrix}
        \mbf r_f \\ \mbf v_f
    \end{bmatrix} = 
    \begin{bmatrix}
        \mathcal M^{[\mbf r],KO}_{\mbf{\bar r}_0, \theta_0 \rightarrow \theta_f}  (\mbf v_0) \\
        \mathcal M^{[\mbf v],KO}_{\mbf{\bar r}_0, \theta_0 \rightarrow \theta_f}  (\mbf v_0)
    \end{bmatrix} 
\end{equation}
The resulting KOSTPM $\mathcal M^{[\mbf r],KO}_{\mbf{\bar r}_0, \theta_0 \rightarrow \theta_f}  (\mbf v_0) $ includes everything that is needed to solve the problem of finding every possible velocity that brings a space object onto the desired final position. This map is square since it has the three inputs, the initial velocity, and the three outputs, the final position, but it is not invertible due to polynomial dependency. In fact, being true anomaly the independent variable, there is an infinite amount of possible $\mbf v_0$ that connect to $\mbf r_f$. Depending on the eccentricity value, there is an infinite amount of orbits spamming the same true anomaly.  

Consequently, the KO solution solution switches from a mere polynomial inversion and evaluation to the problem of finding the zeros of a function. Given the numerical boundary condition of the TPBVP at the final anomaly, $\mbf{\bar r}_f$, and moving it to the left-hand side of the equation holds
\begin{equation}
\mathcal M^{[\mbf r],KO}_{\mbf{\bar r}_0, \theta_0 \rightarrow \theta_f}  (\mbf v_0) - \mbf{\bar r}_f  = \mbf 0 \label{eq_kozeros}
\end{equation}
where any solver available in the state of the art can be used to derive the three-dimensional curve of $\mbf v_0$ that finds the zeros of the equation. Polynomial interpolation and fitting have shown reliable approximation of the null curve. The solution curve is a hyperbola where the two separate branches correspond to the prograde and retrograde solution of the rendezvous problem. 

Thanks to the use of true anomaly as the independent variable, the locus of points where the function is null can be expressed directly as a separate polynomial. In previous works by other authors, this curve is usually represented as an array of floating point numbers, where each point is connected to a specific propagation time. The KOSTPM map, being based on polynomials, directly provides the analytical approximation of this hyperbola (one polynomial for each branch), and subsequently, it is possible to connect any point of the polynomial to its correspondent time of flight \cite{arnas2021approximate}.

%%%%%%%%%%%%%%%%%%%%%%%%%%%%%%%%%%%%%%%%%%%%%%%%%%%%%%%%%%%%%%%%%%%%%%%%%%%%%%%
%%%%%%%%%%%%%%%%%%%%%%%%%%%%%%%%%%%%%%%%%%%%%%%%%%%%%%%%%%%%%%%%%%%%%%%%%%%%%%%
%%%%%%%%%%%%%%%%%%%%%%%%%%%%%%%%%%%%%%%%%%%%%%%%%%%%%%%%%%%%%%%%%%%%%%%%%%%%%%%
\section{The Differential Algebra Solution}
The proposed approach to evaluate the set of velocities hyperbolas is not restricted to the Koopman Operator, but it can be expanded to different methodologies that solve for the family of transfer orbits from different perspectives. The Differential Algebra (DA) implementation has been proven to be a valid solution \cite{armellin2018multiple}, and it provides the tools to obtain the desired outputs. The main idea of DA is to expand the dynamics in their Taylor series expansion around a well-selected center. The expansion order can be freely tuned according to the desired level of accuracy requested by the application, with the drawback of increasing the computational burden on the processing unit. 

The DA solution of a dynamical system uses a numerical propagator to create a polynomial map, based on Taylor polynomials, that directly connects the initial to the final state \cite{servadio2022maximum}. The Taylor polynomial of the deviations around the initial center is propagated forward in time, and the full nonlinear dynamics is approximated. Therefore, the DA solution has been proven beneficial whenever an elevated amount of propagations must be performed, especially for particle filters, where the faster polynomial evaluation of the DA representation of the dynamics substitutes many numerical integrations \cite{servadio2021differential}.

This work exploits another feature of the Differential Algebra Core Engine (DACE2.0) tool, the Taylor polynomial map inversion. That is, having selected the boundaries of a two-point boundary value problem, the inversion of the dynamical map expresses the initial state as a function of the final state. This feature is particularly helpful when solving inverse control problems in the case where the dynamics are expanded using Lagrange multipliers \cite{KOinversejournal,di2008high}. Therefore, the values of the costates can be expressed as a function of the initial and final conditions. Moreover, this methodology is not restricted to the DACE but can be implemented with any map, e.g., the Koopman maps \cite{KOinversejournal}. 

The DA approach is applied directly to the Cartesian equations of motion, with time as the independent variable. 
\begin{equation}
\begin{aligned}
\frac{d x}{d t} &=v_x\\
\frac{d y}{d t} &=v_y\\
\frac{d x}{d t} &=v_z \\
\frac{d v_x}{d t} &=-\dfrac{\mu x}{r^3} +\dfrac{3\mu J_2 R_{\oplus}^2}{2r^5} \bigg( \dfrac{5z^2}{r^2} - 1 \bigg)x\\
\frac{d v_y}{d t} &=-\dfrac{\mu y}{r^3} +\dfrac{3\mu J_2 R_{\oplus}^2}{2r^5} \bigg( \dfrac{5z^2}{r^2} - 1 \bigg)y\\
\frac{d v_z}{d t} &=-\dfrac{\mu z}{r^3} +\dfrac{3\mu J_2 R_{\oplus}^2}{2r^5} \bigg( \dfrac{5z^2}{r^2} - 3 \bigg)z
\end{aligned}
\label{eqn:cartesianj2}
\end{equation}
where $\mbf r = [x, y, z]^T $ and $\mbf v = [v_x, v_y, v_z]^T$ are the spacecraft position and velocity vectors. Since time is an independent variable, the DA solution of the two-point boundary value problem (TPBVP) has the boundary position vectors connected by a specific time of flight, defined as the time between the initial condition, $t_0$, and the final time step, $t_f$. Therefore, thanks to the propagation of the Taylor polynomial expansion series, the solution is in the form of a Differential Algebra State Transition Polynomial Map (DASTPM), which can be written as
\begin{equation}
    \mbf x_{f,C} = \mathcal M^{DA}_{t_0 \rightarrow t_f}  (\mbf x_{0,C}) \label{DAmap}
\end{equation}
When comparing this formulation with the KO map derived in Eq. \eqref{KOmap}, it can be noted that the main difference relies on the different independent variable. The KOSTPM represents all orbits that span the same change in true anomaly, while the DASTPM represents all orbits that are propagated for the selected time of flight.

%%%%%%%%%%%%%%%%%%%%%%%%%%%%%%%%%%%%%%%%%%%%%%%%%%%%%%%%%%%%%%%%%%%%%%%%%%%%%%%
\subsection{The DASTPM Velocity Hyperbola }
This representation of the dynamics is exploited to evaluate the curve of velocities that connect two position vectors in space. The polynomial map can be deconstructed into the position and velocity part halves. The first half highlights the dependency of the final position vectors on the initial position and velocity vectors. Thus, the map ensures a connection between the two positions. This is similar to Lambert's problem but generalized to include any possible velocity solution and accounting for the perturbation terms. 

The DASTPM expressed by Eq. \eqref{DAmap} connects the full initial state to the final state at $t_f$. Since the focus of the methodology is to connect position vectors working on the velocity, the six-dimensional map can be split into two parts: 
\begin{equation}
    \mbf x_{f,C} = 
    \begin{bmatrix}
        \mbf r_f \\ \mbf v_f
    \end{bmatrix} = 
    \begin{bmatrix}
        \mathcal M^{[\mbf r],DA}_{t_0 \rightarrow t_f}  (\mbf x_{0,C}) \\
        \mathcal M^{[\mbf v],DA}_{t_0 \rightarrow t_f}  (\mbf x_{0,C}) 
    \end{bmatrix} = 
    \begin{bmatrix}
        \mathcal M^{[\mbf r],DA}_{t_0 \rightarrow t_f}  (\mbf r_0, \mbf v_0) \\
        \mathcal M^{[\mbf v],DA}_{t_0 \rightarrow t_f}  (\mbf r_0, \mbf v_0) 
    \end{bmatrix}
\end{equation}
where the dependency on the initial position $\mbf r_0$ and initial velocity $\mbf v_0$ have been marked. Only the first half of the DASTPM is of interest when dealing with debris threat evaluation, as it gives information regarding the trajectory of the debris, while the second half informs regarding the velocity at which the debris is traveling. When designing avoidance maneuvers, it is not necessary to know the velocity of the debris if the asset spacecraft never crosses its pathway. 

The location of the collision event, $\mbf{\bar r}_0$, which indicates the initial position vector, is known. For clarity, $\mbf{\bar r}_0$ is the actual numerical value of the boundary of the TPBVP, while $\mbf r_0$ is the initial position variable. The position DASTPM can be evaluated at the given vector to remove the dependency on the position variables, leaving only the influence on the initial velocity vector:
\begin{equation}
 \mbf r_f = \mathcal M^{[\mbf r],DA}_{t_0 \rightarrow t_f}  (\mbf{\bar r}_0, \mbf v_0) = 
 \mathcal M^{[\mbf r],DA}_{\mbf{\bar r}_0, t_0 \rightarrow t_f}  (\mbf v_0) \label{DAmapfinal}
\end{equation}
This expression of the polynomial map shows that there is dependency only on the initial velocity and that the map is centered at $\mbf{\bar r}_0$. The position DASTMP centered at the given initial condition defined in Eq. \eqref{DAmapfinal} is square: it takes three input variables, the initial velocity components, and outputs three results, the final position components. Thus, having independent polynomials due to the Cartesian definition of the dynamics, this map can be inverted. The map inversion is performed directly in the DA framework and stands
\begin{equation}
 \mbf v_0 = \Big( 
 \mathcal M^{[\mbf r],DA}_{\mbf{\bar r}_0, t_0 \rightarrow t_f}  (\mbf v_0) \Big)^{-1} = 
 \mathcal W^{[\mbf r],DA}_{\mbf{\bar r}_0, t_f \rightarrow t_0}  (\mbf r_f) \label{DAinv}
\end{equation}
where the input and output have been switched so that the inverted map $\mathcal W$ describes the values of the initial velocity as a function of the final position. That is, Eq. \eqref{DAinv} is a 1-on-1 function that identifies the requested initial velocity of the spacecraft to select an orbit that will reach the desired final point given a flight time. Therefore, the numerical value of the initial velocity is calculated through polynomial evaluation of the inverted DASTPM, which is just a mere polynomial substitution of variables. Indeed, given the numerical value of the desired final position, $\mbf{\bar r}_f$, the initial velocity that corresponds to that orbit is 
\begin{equation}
 \mbf{\bar v}_0 = 
 \mathcal W^{[\mbf r],DA}_{\mbf{\bar r}_0, t_f \rightarrow t_0}  (\mbf{\bar r}_f) 
\end{equation}
The vector $\mbf{\bar v}_0$ is just a single point of the hyperbola of velocities that connect $\mbf{\bar r}_0$ to $\mbf{\bar r}_f$, corresponding to the time of flight $t_f - t_0$. The whole curve of velocities can be outlined by varying the time of flight and repeating the process. 

The problem is solved, and the DA solution has been found, given the TPBVP conditions, the set of orbital velocities that the debris must have to intersect a point in space. Unlike the Koopman Operator approach, where the velocity hyperbola was directly the solution of the KOSTPM, the DA approach evaluates each point of the curve separately, creating a map for each flight time. Regardless, both maps account for the $J_2$ perturbation term, covering the same change in true anomaly. 

It is important to underline some differences between the two approaches. The KO solution of the dynamics is a global representation of the equations of motion, while the DA approach provides a local approximation based on Taylor polynomials. While KO learns the coefficients of its map by projecting the dynamics onto the whole set of orthogonal basis functions for the whole domain, the DA solution creates its map by expanding the polynomial around a well-selected center. Lastly, the different independent variable selected by the two approaches represents the main reason behind a point-wise solution for the DA approach, where the velocity hyperbola is evaluated point-by-point by changing the time of flight, in comparison to a curve of zeros for the KO approach where the hyperbola is the locus of points where the map has a value of zero. This aspect is appreciated by looking at the definition of the span in true anomaly of an orbit between two points:
\begin{equation}
    \cos\Delta\theta = \dfrac{\mbf{\bar r}_0 \cdot \mbf{\bar r}_f }{||\mbf{\bar r}_0 ||\ ||\mbf{\bar r}_f ||}
\end{equation}
since there is no time dependency in the equation. This feature is helpful when it is desired to study multiple revolution solutions \cite{pasiecznik2022lambert}. Indeed, in the DA framework, the velocity hyperbola for multiple revolutions orbit requires particular care \cite{armellin2018multiple} since if it were to increase the time of flight, the DA solution will only provide a longer approximation of the initial hyperbola and not a new one. On the contrary, using the KOSTPM, it is possible to include multiple revolutions by adding a factor of $2\pi$ when creating the map, obtaining a final equation to solve in the form
\begin{equation}
\mathcal M^{[\mbf r],KO}_{\mbf{\bar r}_0, \theta_0 \rightarrow 2\pi \zeta+\theta_f}  (\mbf v_0) - \mbf{\bar r}_f  = \mbf 0 \label{eq:multi}
\end{equation}
where $\zeta$ is the number of revolutions. The whole KO process is untouched, and nothing has changed except the selected span of true anomaly. 

However, the DA solution is easier to implement and more intuitive to understand, as it works directly on the not transformed dynamics without the need for any mathematical tricks to obtain a smoother approximation, likewise the Koopman Operator. 

%%%%%%%%%%%%%%%%%%%%%%%%%%%%%%%%%%%%%%%%%%%%%%%%%%%%%%%%%%%%%%%%%%%%%%%%%%%%%%%
%%%%%%%%%%%%%%%%%%%%%%%%%%%%%%%%%%%%%%%%%%%%%%%%%%%%%%%%%%%%%%%%%%%%%%%%%%%%%%%
%%%%%%%%%%%%%%%%%%%%%%%%%%%%%%%%%%%%%%%%%%%%%%%%%%%%%%%%%%%%%%%%%%%%%%%%%%%%%%%

\section{The NASA Standard Breakup Model}
A commonly used breakup model for on-orbit collisions and explosion model is the NASA Standard Breakup Model (SBM) \cite{nasaSTMevolve4}, a data-driven sampling-based model developed from data collected from on-orbit explosions, collisions, and several ground-based explosions and hyper-velocity collision experiments. Given two colliding objects, the SBM outputs the size, area-to-mass ratio, and imparted $\Delta V$ for each generated fragment. Collisions are defined into catastrophic and non-catastrophic events, depending on the specific kinetic energy threshold of 40 J/g. The cumulative number of fragments for some given minimum-sized debris follows a power law. 

As originally published, the SBM does not guarantee mass conservation; however, this paper will guarantee the conservation of mass and momentum. At the instant of the collision, the SBM assumes a distribution of imparted velocities as spherically symmetrical around each of the objects that fragment. Fragmentation events are also simulated with the NASA break-up model \cite{nasaSTMevolve4, KlinkradBook2006}. It starts with defining the characteristic length, $\ell_c$, which is considered constant and equal to 0.05 m. Note that $\ell_c$ will be treated as an equivalent diameter, $d$. The number of fragments $n_f$ of diameter $d>\ell_c$ can be computed as 
\begin{equation}
    n_f = \begin{cases}
    6\, c_s \, \hat{\ell}_c^{-1.6} &\quad \textrm{for explosions}\\
    0.1\, \hat{m}^{0.75} \, \hat{\ell}_c^{-1.71} &\quad \textrm{for collisions}
    \end{cases}
    \label{eq:NASAsbm}
 \end{equation}
where $\hat{\ell}_c = {\ell}_c / [\textrm{m}]$ and 
\begin{align}
    \hat{m} = \begin{cases}
    \dfrac{m_t+m_p}{\textrm{[kg]}} &\quad \textrm{for } \Tilde{E}_p\geq\Tilde{E}_p^*\\
    \dfrac{m_p\,v_i^2}{\textrm{1000[kg\,(m/s)$^2$]}} &\quad \textrm{for } \Tilde{E}_p<\Tilde{E}_p^*
    \end{cases} && \quad \textrm{with} &&
    \Tilde{E}_p = \frac{m_p\,v_i^2}{2\,m_t}
 \end{align}
Note that the symbol $\,\hat{}\,$ indicates normalized quantities, $m_t$ and $m_p$ are respectively the target (mostly derelicts or rocket bodies) and projectile mass, $v_i$ the impact velocity (considered constant and equal to 10 km/s), $\Tilde{E}_p$ the specific energy of the projectile, and $\Tilde{E}_p^*=40$ [kJ/kg] the specific energy threshold for a catastrophic collision. Particular attention should be paid to the scaling parameter $c_s$ in Eq. \eqref{eq:NASAsbm}. It is an event-specific calibration constant for historic events and an empirical correction for certain classes of fragmentation events in the case of future projections (with $0.1 \leq c_s \leq 1.0$). For masses between 600 kg and 1000 kg, the calibration factor is $c_s = 1.0$. However, past fragmentation events showed very different characteristics; thus, the break-up models must be calibrated. For this procedure, let us indicate with $d_{tr}$ the trackable diameter computed from the Tracking Relevant Altitude (TRA), $H$, which is assumed to be the perigee altitude of the parent object $h_p$:
\begin{equation}
    H = h_p
\end{equation}
To determine the trackable diameter, an empirical formula is used
\begin{equation}
    d_{tr}=\begin{cases}
    8.9 \textrm{ cm}  & \quad H\leq 620 \textrm{ km} \\
    1.0 \textrm{ cm}\cdot 10^{-0.736748+0.604~\log(H)}   & \quad 620\textrm{ km} < H \leq 1300\textrm{ km} \\
    1.0 \textrm{ cm}\cdot 10^{-4.417+1.8186~\log(H)}   & \quad 1300\textrm{ km} < H \leq 3800\textrm{ km} \\
    1.0 \textrm{ m}& \quad H > 3800 \textrm{ km} \\
    \end{cases}
\end{equation}
The number of objects larger than the trackable diameter is computed by applying a correction through the Henize factor $f_{Hz}$, which is computed as 
\begin{equation}
    f_{Hz} = \begin{cases}
    \sqrt{ 10^{e^{\left(-\frac{\log{(\hat{d}_{tr})} - 0.78}{0.637}\right)^2}} } & \quad \textrm{if } d_{tr}>10^{0.78} \textrm{ cm}\\
    \sqrt{10}  & \quad \textrm{otherwise} 
    \end{cases}
\end{equation}
The resulting value $n_f$ can now be considered as the true number of objects larger than $d_{tr}$ generated by an event on an orbit with perigee altitude $H$.

According to the NASA standard break-up model, the area-to-mass ratio $A/m$ for new fragments is assigned according to a bi-modal probability density function $p(\chi, \vartheta)$.
\begin{equation}
    p(\chi, \vartheta) = \alpha(\vartheta)~p_1(\chi) + (1-\alpha(\vartheta))~p_2(\chi)
\end{equation}
where $\chi = \log_{10}(\{A/m\}/[\textrm{m}^2/\textrm{kg}])$ is the area-to-mass parameter, $\vartheta=log_{10}(d/[\textrm{m}])$, and $p_{1,2}$ indicates the normally distributed density functions. The parameter $\alpha$, the means $\mu_{1,2}$, and standard deviations $\sigma_{1,2}$ are computed as stated in the NASA's new break-up model of EVOLVE 4.0 \cite{nasaSTMevolve4}. According to this model, the effective cross-section $A$, function of the fragment diameter $d$, is
\begin{equation}
    A/[\textrm{m}^2]= \begin{cases}
    0.540424 (d/[\rm m])^2 & \quad \textrm{for}~d<1.67 \textrm{mm}\\
    0.556945 (d/[\rm m])^{2.0047077} & \quad \textrm{for}~d\geq1.67 \textrm{mm}\\
    \end{cases}
\end{equation}
The fragment mass is thus determined as
\begin{equation}
    m = \dfrac{A}{A/m}
\end{equation}
The model also requires assigning the imparted fragmentation velocities, which are sampled from a normal distribution characterized by the following mean value and standard deviation
\begin{equation}
  \begin{aligned}
    \mu_\nu = 0.2 \log_{10}(A/m) + 1.85 && \sigma_\nu=0.4 &&& \textrm{for explosions}\\
    \mu_\nu = 0.9 \log_{10}(A/m) + 2.90 && \sigma_\nu=0.4 &&& \textrm{for collisions}
  \end{aligned}
\end{equation}
where $\nu = \log_{10}(\Delta v)$. These velocities are the paper's main focus, as they describe the future trajectories of the debris, which might intersect the orbit of the asset. Therefore, an analysis of the distribution of these velocities depending on the type of collision is performed, and the resulting distribution is compared to the family of velocities that represent a threat to the asset. Moreover, due to the stochastic nature of the model, statistical analyses must be performed to ensure an accurate representation of the families of debris generated by a break-up event in space.

%%%%%%%%%%%%%%%%%%%%%%%%%%%%%%%%%%%%%%%%%%%%%%%%%%%%%%%%%%%%%%%%%%%%%%%%%%%%%%%
%%%%%%%%%%%%%%%%%%%%%%%%%%%%%%%%%%%%%%%%%%%%%%%%%%%%%%%%%%%%%%%%%%%%%%%%%%%%%%%
%%%%%%%%%%%%%%%%%%%%%%%%%%%%%%%%%%%%%%%%%%%%%%%%%%%%%%%%%%%%%%%%%%%%%%%%%%%%%%%
\section{Convex Hull of Debris}
The NASA SBM creates the set of possible velocities that debris might have after a  breakup event in space. Due to the stochastic nature of the event, each time a collision is simulated, a different set of velocities is created, meaning that, in order to have a comprehensive representation of the debris, the simulation must be repeated multiple times, following a Monte Carlo approach. 

This procedure results in a list of orbits represented according to their velocity vectors since they share the same position, which is the collision or explosion location. However, regardless of how many simulations are performed, the point-wise representation of the set of debris velocities is incomplete and faulty since any velocity between two outcomes could have been a valid solution. The SBM provides a discrete representation of velocities over the desired continuous function. Therefore, it is needed to implement techniques that encapsulate the region of space in the velocity domain that encloses all of the possible velocities that debris could have after the breakup event. The QuickHull algorithm implemented in \cite{barber1996quickhull} has been chosen to address this problem. 

\begin{figure}[ht]
    \centering
    \includegraphics[width=0.45\textwidth]{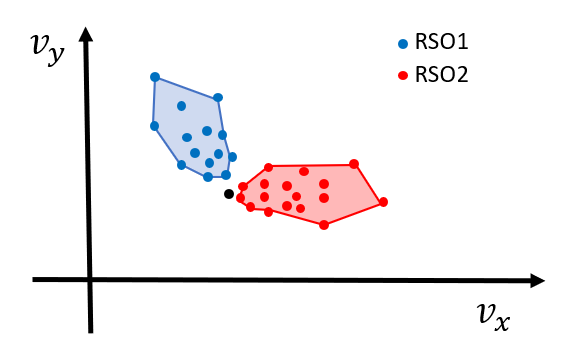}
    \caption{Convex Hull representation for a collision event in 2D.}
    \label{fighull}
\end{figure}

Figure \ref{fighull} gives a visual representation, in two dimensions, of the idea behind the formation of the convex hull. After a collision is simulated at a specific point, the SBM provides the changes of velocity, likewise impulses, for the debris resulting from the breakup event. In the figure, the velocities of the debris related to the first colliding resident space object, RSO1, and the second, RSO2, are reported with blue and red points, respectively. The algorithm creates the convex hull of the set of points, which is the smallest convex set that contains all of the points. Therefore, any velocity inside the hull could have resulted from the breakup event. In other words, the convex hulls, red and blue shaded areas in the figure, are the closed convex regions that enclose all the possible velocities the debris could have. 

After an elevated amount of SBM simulations, the convex hull represents, with high fidelity, the velocity domain where a debris' velocity vector could lay. However, it gives no information regarding the likelihood inside the hull. As expected, more points are close to the original velocity (before the collision), and fewer points are located far from it. Thus, the probability of the hull is not uniform, but the density of points decreases as the region gets farther from the collision event. The hull represents the velocity probability density function (PDF) domain. The actual PDF curve can be inferred and interpolated directly by the points; however, since the procedure aims to guarantee safety from any possible generated debris, the knowledge of the hull is enough information to perform a safety evaluation. In fact, if the velocity hyperbolas previously described never intersect the hull, there is no worry that debris could hit the asset under protection. 

\begin{figure}[ht]
    \centering
    \includegraphics[width=0.8\textwidth]{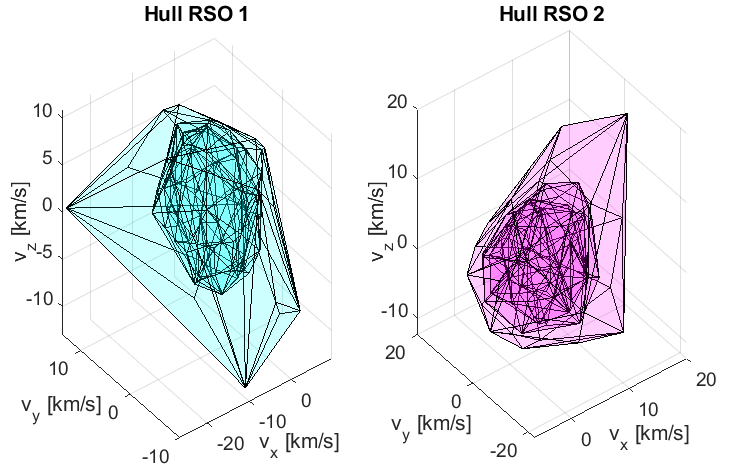}
    \caption{Convex Hull for RSO 1 and RSO 2.}
    \label{fighull2}
\end{figure}

As the particles are densely concentrated around their mean, with a few spreading far from the rest of the debris, a problem connected to the definition of a PDF domain arises. For example, consider a one-dimensional Gaussian distribution with a given mean and covariance. Even if the domain of the PDF is the whole set of Real numbers, since the Gaussian bell goes to zero at infinity, almost all of the distribution is enclosed inside a boundary of five times the standard deviation. It is then safe to assume that the Gaussian distribution has a finite domain with a length of 10 standard deviations, even if it can happen to have an outcome particularly far from the mean. Similarly, some convex hull points might stretch the region and considerably increase the hull volume. These points have extremely low probability and result from the elevated number of Monte Carlo simulations performed. Therefore, a recursive algorithm for removing extremely low probability points has been implemented to have a more realistic representation of the hull that does not stretch too much outside its actual feasible domain. This selection has been made to obtain a realistic convex hull. The QuickHull algorithm selects the outer points that define the convex hull, and its volume is easily calculated using geometry, $V_0$. The hull is then reevaluated without considering these points. The result is a smaller convex hull inside the previous one, with volume $V_1$. After defining a threshold $\tau$, which represents the relative ratio of how much the volume between two consecutive hulls is reduced, the process of removing the most outer points of the hull is stopped when  
\begin{equation}
    \dfrac{V_{i-1}-V_i}{V_i} < \tau
\end{equation}
where $i$ indicates the iteration number. This procedure ensures that the final hull is less dependent on the randomness of the Monte Carlo generation of debris according to the SBM, and it avoids unrealistic large hulls where one debris is so far from the rest of the other points that it increases the volume of the hull to unrealistic boundaries. Figure \ref{fighull2} shows the results of the two convex hulls after undergoing their reduction iterations, with a value of $\tau = 0.5$, which means that, in one iteration, two consecutive hulls lost more than half of their volume. The figure is obtained after merging together all the debris generated by $10^5$ runs of the SBM for a single collision, resulting in more than $10^8$ debris resolutions. The figure shows that three separate hulls are evaluated thanks to the recursive use of the QuickHull algorithm ($i = 2$), keeping the inner one for threat evaluation analysis. 
The Hull for RSO 1 in Figure \ref{fighull2} clearly shows how two points out of $10^8$ stretched the initial hull far from the high-density region. After the second iteration, the resulting hull better describes the real debris distribution and the debris's domain PDF. Without the iterative check on the volume, two points would have compromised the outcome of the analysis and created a bias in the results. 

Regardless of how accurate a model can be, the SBM cannot represent the true collision behavior, and some assumptions must be made. Looking at Fig. \ref{fighull2}, it can noted how the hull covers the full velocity domain, with changes in velocities that span up to 30 km/s, being unrealistic. If the threat evaluation were performed with this definition of the hull, no asset orbit would be considered safe, as there would always be an intersection between the hull and the velocity hyperbolas. This issue comes from the mass conservation feature of the SBM, where particularly small debris, in the order of a few grams, are shot at extremely high speeds to ensure the system's energy conservation principle. For example, the output from the SBM reported in the figure has a single debris with a mass of 12 g traveling at more than 30 km/s. Therefore, in order to represent truthful hulls that represent the actual collision behavior, debris under the mass of 1 kg have been neglected.

%%%%%%%%%%%%%%%%%%%%%%%%%%%%%%%%%%%%%%%%%%%%%%%%%%%%%%%%%%%%%%%%%%%%%%%%%%%%%%%
%%%%%%%%%%%%%%%%%%%%%%%%%%%%%%%%%%%%%%%%%%%%%%%%%%%%%%%%%%%%%%%%%%%%%%%%%%%%%%%
%%%%%%%%%%%%%%%%%%%%%%%%%%%%%%%%%%%%%%%%%%%%%%%%%%%%%%%%%%%%%%%%%%%%%%%%%%%%%%%
\section{The Koopman Operator Optimization}
The KO techniques described in the previous section can be applied to solve orbit transfer optimization problems. After all, the KO solution of the velocity hyperbola provides information on every possible transfer orbit connecting two points in space, and therefore, having the analytical function, minimum distances and derivatives are easily calculated. This aspect can be appreciated in the proposed application. 

%%%%%%%%%%%%%%%%%%%%%%%%%%%%%%%%%%%%%%%%%%%%%%%%%%%%%%%%%%%%%%%%%%%%%%%%%%%%%%%
\subsection{Energy Optimal}
\begin{figure}[!htb]%
\centering
\subfigure[Transfer Orbit Family.]{%
\label{figfamilyorbit}%
\includegraphics[width=.5\linewidth]{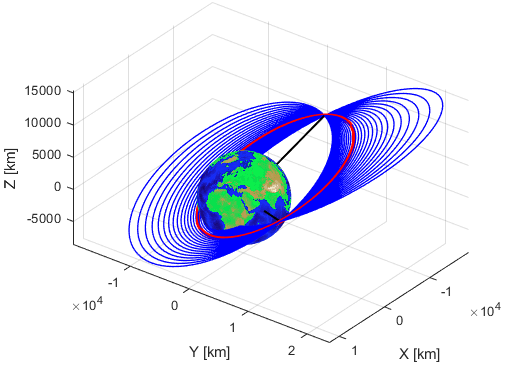}}%
\subfigure[Velocity Hyperbola.]{%
\label{velohyp}%
\includegraphics[width=.5\linewidth]{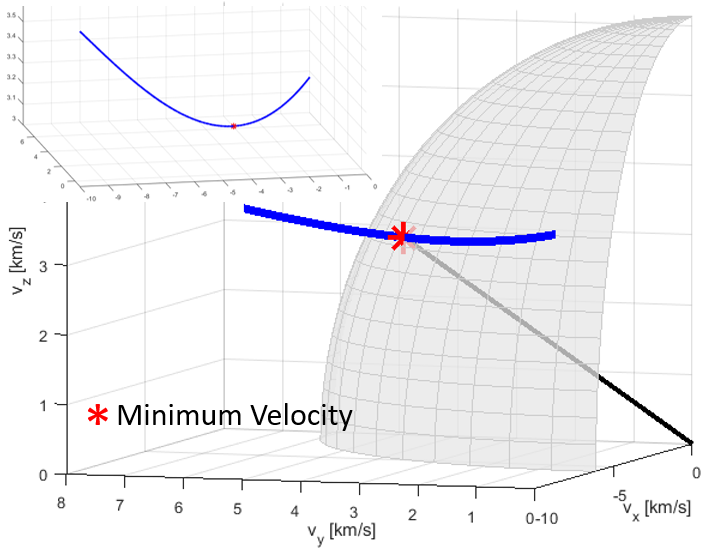}}%
\caption{Minimum Energy Solution among every possible transfer orbit.}
\label{fig:energyopt}  
\end{figure}

Following Lambert's problem application in \cite{pasiecznik2022lambert}, it is desired to derive the family of orbits around Earth that connect the two points in the Earth Central Inertial (ECI) reference frame:
\begin{align}
    \mbf r_0 &= [5000\quad 10000\quad 2100]^T \text{  km} \\
    \mbf r_f &= [-14600\quad 2500\quad 7000]^T \text{  km}
\end{align}
Using the KOSTPM solution of the dynamics, the family of orbits that pass through both points is uniquely evaluated. As the relation is a 1-on-1 function, at each orbit corresponds a specific change in velocity from the collision event. By working with true anomaly as the independent variable, when solving Eq. \eqref{eq_kozeros}, the KOSTPM finds the curve, expressed as interpolated polynomials, in the velocity space of orbits that are a candidate for Lambert's problem. The curve in the velocity domain and the relative family of orbits are reported in blue in Fig. \ref{fig:energyopt}, respectively. Every solution is propagated using the same Koopman matrix that connects the position vectors in the TPBVP, as the change in the anomaly of the transfer orbit is the same regardless of the transfer time. That is, looking at the family of transfer orbits, Fig. \ref{figfamilyorbit}, they all have a different semi-major axis, a different time of flight, and a different velocity vector, but they all cover the same true anomaly span, $\Delta \theta$. Therefore, for a specific point in the spacecraft orbit, the family of velocities that create orbits connecting the two points is evaluated in the KO framework. This curve, called $\Xi(\theta)$, is represented in the three-dimensional velocity space $\{\dot x,\dot y,\dot z\}$ in Figure \ref{velohyp}.

The analytical function of the curve suits the proposed methodology for the application of various optimization techniques and analysis. For example, since each orbit starts from the same position vector, i.e., constant range $r$, the orbit transfer with the minimum energy is the one with the minimum velocity since the specific energy of an orbit is defined as
\begin{equation}
    E = \dfrac{v^2}{2} - \dfrac{\mu}{r} = -\dfrac{\mu}{2a} \label{eq:energy}
\end{equation}
where $v$ is the velocity vector magnitude and $a$ is the semi-major axis of the transfer orbit. The orbit with the minimum energy is connected to the hyperbola's shortest velocity vector. Figure \ref{velohyp} highlights the point of $\Xi(\theta)$ closest to the origin, which corresponds to the minimum velocity norm and, therefore, to the minimum energy transfer. The transfer is highlighted in red in Fig. \ref{figfamilyorbit}. Equation \eqref{eq:energy}, at the same time, identifies the smallest orbit with the lowest semi-major axis since it is the one with the lowest specific energy value. 

%%%%%%%%%%%%%%%%%%%%%%%%%%%%%%%%%%%%%%%%%%%%%%%%%%%%%%%%%%%%%%%%%%%%%%%%%%%%%%%
\subsection{Fuel Optimal}
\begin{figure}[!htb]%
\centering
\subfigure[Minimum Fuel Transfer.]{%
\label{figfamilyfuel}%
\includegraphics[width=.5\linewidth]{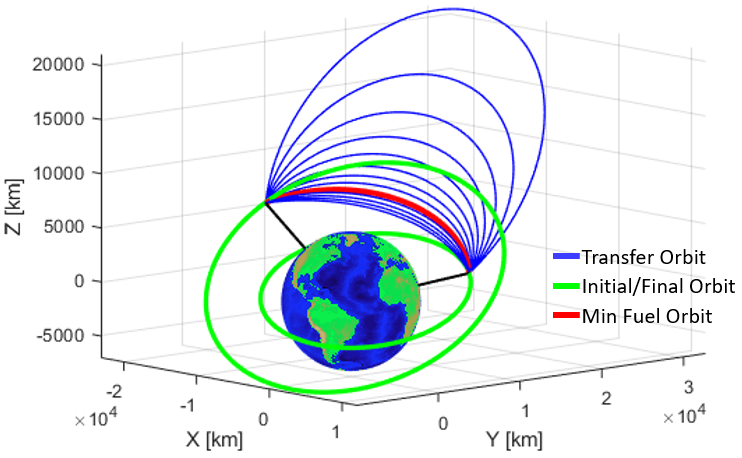}}%
\subfigure[Velocity Hyperbolas.]{%
\label{velofuel}%
\includegraphics[width=.5\linewidth]{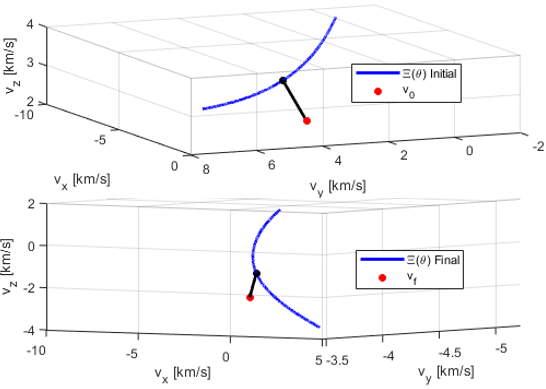}}%
\caption{Minimum Fuel Solution among every possible transfer orbit.}
\label{fig:fuelopt}  
\end{figure}

The family of possible transfer orbits has been parameterized for each point of curve $\Xi(\theta)$. This representation leads to other kinds of optimizations, such as minimum fuel transfer, with the assumption of impulsive maneuvers. Thus, assume that the spacecraft is moving between two separate orbits rather than two points in space. The initial and final orbits are defined by selecting velocity vectors to complete the state of the two-body problem system:
\begin{align}
    \mbf v_0 &= [-4.3\quad +2.6\quad +2.0]^T \text{  km/s} \\
    \mbf v_f &= [-1.0\quad -3.8\quad -2.3]^T \text{  km/s}
\end{align}
The two orbits are reported in green in Fig. \ref{figfamilyfuel}. Knowing the velocity vector at the giving point and the family of transfer orbit $\Xi(\theta)$, the minimum fuel solution corresponds to the transfer orbit that requires the minimum total impulse to let the spacecraft change from initial to transfer orbit and from transfer to final orbit. Figure \ref{figfamilyfuel} reports in red the minimum fuel solution, while Fig. \ref{velofuel} shows the impulse vectors in the velocity space as the difference between the velocity hyperbolas at the initial and final conditions. 

The minimum condition is obtained with a global search on the sum of two impulses to identify the transfer orbit that requires the smallest overall change in velocity. In the selected application, the smallest total impulse is $\Delta V_{MIN} = 2.480$ km/s, composed by a first impulse of 1.356 km/s and a second of 1.125 km/s. Knowing the velocity hyperbola at the initial and final position, thanks to the KOSTPM leads to an easy scalar optimization problem among vectors' magnitudes. The minimum solution, $\Delta V_{MIN}$ of the scalar $\Delta V_{TOT}$ function is shown in Fig. \ref{deltavtot} with a red point.

%%%%%%%%%%%%%%%%%%%%%%%%%%%%%%%%%%%%%%%%%%%%%%%%%%%%%%%%%%%%%%%%%%%%%%%%%%%%%%%
\subsection{Time Optimal}
\begin{figure}[!htb]%
\centering
\subfigure[Minimum Transfer Time.]{%
\label{figfamilytime}%
\includegraphics[width=.5\linewidth]{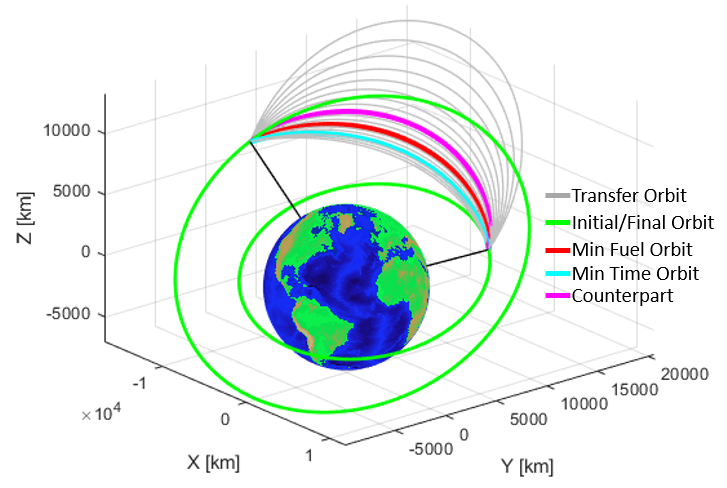}}%
\subfigure[Total Impulse Curve.]{%
\label{deltavtot}%
\includegraphics[width=.5\linewidth]{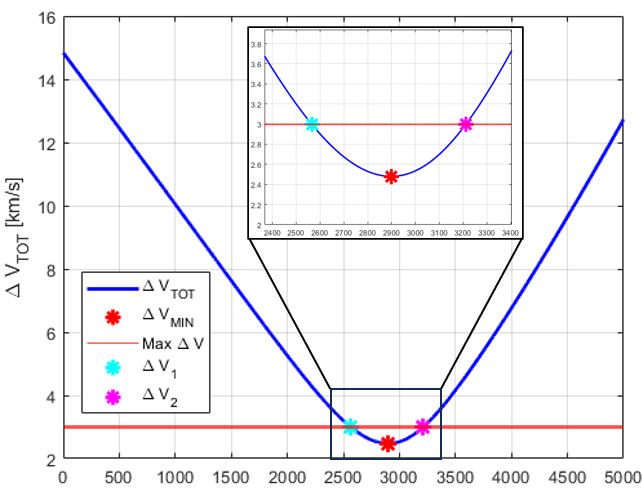}}%
\caption{Minimum Time-Of-Flight solution among every possible transfer orbit.}
\label{fig:timeopt}  
\end{figure}

Lastly, time optimization can be performed under the assumption that a maximum amount of fuel is available. Assuming a maximum total impulse of $\Delta V_{MAX} = 3$ km/s, the minimum time of flight optimization is translated to determining the transfer orbit with a total impulse equal to the constraint, as that corresponds to the fastest transfer orbit. That is, all the fuel is used to achieve the maximum change in velocity to set the spacecraft to the fastest route. Figure \ref{deltavtot} shows the maximum impulse constraint on the total impulse curve as a red horizontal line. The constraint crosses the function in two separate points, $\Delta V_1$ and $\Delta V_2$. One point corresponds to the minimum time-of-flight solution, while the other is its counterpart, i.e., the orbit with the same change in velocity but selected on the opposite side with respect to the minimum impulse orbit. The solution pathway is reported in Fig. \ref{figfamilytime}, where the light blue orbit is the optimal time orbit, while the red orbit is the minimum impulse one from the previous analysis. The magenta orbit is the counterpart. 

The analytical parameterization of every possible transfer orbit between two orbits has been proven useful for optimization applications. Regardless of the type of optimization required, the approximation of the velocity hyperbola from the KOSTPM using polynomials leads to an easy and fast calculation of values on the velocity domain. It is now possible to elaborate on these results and expand to the arbitrary selection of any true anomaly at the final orbit, covering the collision threat evaluation problem. 

%%%%%%%%%%%%%%%%%%%%%%%%%%%%%%%%%%%%%%%%%%%%%%%%%%%%%%%%%%%%%%%%%%%%%%%%%%%%%%%
%%%%%%%%%%%%%%%%%%%%%%%%%%%%%%%%%%%%%%%%%%%%%%%%%%%%%%%%%%%%%%%%%%%%%%%%%%%%%%%
%%%%%%%%%%%%%%%%%%%%%%%%%%%%%%%%%%%%%%%%%%%%%%%%%%%%%%%%%%%%%%%%%%%%%%%%%%%%%%%
\section{The Cosmos-Iridium Collision Application}
% general cosmos iridium info and data
% DAN TO FILL
On February 10, 2009, at 16:56 UTC,  an inactive Russian military satellite, Kosmos-2251, collided with a functioning commercial satellite, Iridium-33. The collision took place approximately at 800 km and marked the first instance of a collision between two cataloged orbital satellites.  

The impact generated a substantial amount of debris, estimated to be tens of thousands of fragments, adding to the existing space debris in Earth's orbit.  The increasing number of debris from the collisions were tracked and added to the catalog with time. By 2009 Aug 26, the SSN had cataloged 406 debris; by 2012 Aug 31, 2199 debris had been cataloged, where 1602 debris were attributed to Cosmos-2251 and 597 were attributed to Iridium-33. The satellite parameters of the two Resident Space Objects (RSOs) are reported in Table \ref{tab:iridiumcosmosParam}.

\begin{table}
    \centering
    \begin{tabular}{ccccc}
                    & SATCAT ID & COSPAR ID & Mass (kg) & Dimension (m)  \\ \hline 
        Iridium 33  & 24946 & 1997-051C & 556 & 3.1 $\times$ 2.4 $\times$ 1.5  \\  \hline
        Kosmos 2251 & 22675 & 	1993-036A & 900 & 1.5 (radius)
    \end{tabular}
    \caption{Iridium and Kosmos satellite parameters at the time of collision}
    \label{tab:iridiumcosmosParam}
\end{table}

The two RSOs' orbit parameters are reported in Table \ref{tab:iridiumcosmosORBParam}, taken from the catalog TLE right before the collision event. The orbits have two points of close conjuctions, where the trajectories ``intersect". Indeed, the collision event happened at 
\begin{align}
    h &= 788.68 \text{ km} \nonumber \\
    \phi &= 75.5011 \text{ deg} \nonumber \\
          \lambda &= 97.8794 \text{ deg} \nonumber 
\end{align}
where $h$ is the altitude, $\phi$ is the latitude, and $\lambda$ is the longitude of the collision point\footnote{https://web.archive.org/web/20090216151141/http://n2yo.com/collision-between-two-satellites.php}. The possibility of a break-up event was monitored by many SSA software, which highlighted this particular close conjunction as likely to escalate to a collision. However, this conjunction was not the one with the highest probability of collision happening that day, supporting the necessity of threat estimation and safety. 

\begin{table}
    \centering
    \begin{tabular}{cccccc}
                    & $a$ (m)   & $e$       & $i$ (deg) & $\Omega$ (deg) & $\omega$ \\ \hline
        Iridium 33  & 7152.2009 & 0.0002253 & 86.3989   & 121.2960       & 89.6115  \\  \hline
        Kosmos 2251 & 7162.4744 & 0.001615  & 74.0357   & 17.1729        & 95.9865
    \end{tabular}
    \caption{Iridium and Kosmos orbit parameters at the time of collision}
    \label{tab:iridiumcosmosORBParam}
\end{table}

\begin{figure}
    \centering
    \includegraphics[width=0.4\textwidth]{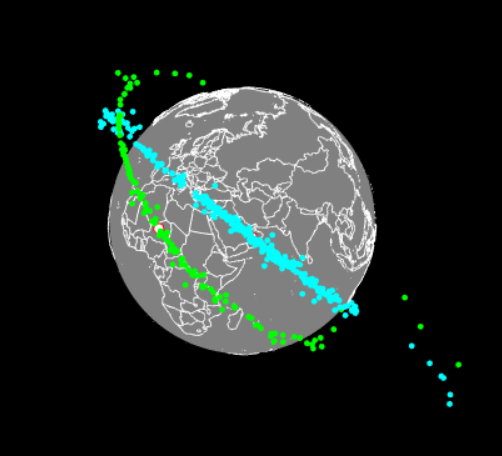}
    \caption{Collision between two satellites after a few revolutions as simulated by MOCAT-MC using the NASA EVOLVE 4.0 parameters}
    \label{fig:collisiongif}
\end{figure}

% https://en.wikipedia.org/wiki/Kosmos_2251
% https://en.wikipedia.org/wiki/Iridium_33
% https://conference.sdo.esoc.esa.int/proceedings/sdc6/paper/45/SDC6-paper45.pdf
% https://www.agi.com/getmedia/d4bbbff7-2e79-48e8-a3ac-2407afff9951/Space-Surveillance-Lessons-Learned-From-The-Iridium-Cosmos-Collision.pdf?ext=.pdf)
Figure \ref{fig:collisiongif} gives a representation of the spread of the debris around the earth after the impact between the two RSOs. According to the different velocity profiles, the debris form a cloud of objects that occupies the whole original orbit. As a consequence of the impact, multiple payloads are forced to implement weekly debris avoidance maneuvers. The figure shows the division of the debris into two separate families depending on their original parent satellite. Indeed, the NASA SBM creates two separate ensembles of debris that lead to two separate hulls since debris behave similarly to either one or the other original RSOs. 

The proposed application uses the Iridium-Kosmos break-up event as if it were to happen today. Therefore, a Starlink Phase 1 payload has been selected as the asset satellite, with orbit parameters reported in Table \ref{tab:starlinkORBParam} according to the most current TLE. The goal of the proposed application is to assess if the selected collision would affect the orbit of the asset, enforcing it to prepare evasive maneuvers. The asset is considered in danger if there is overlapping between the hulls from the SBM and any velocity hyperbolas connecting the break-up location to the asset's orbit. 

\begin{table}
    \centering
    \begin{tabular}{cccccc}
                         & $a$ (m)   & $e$        & $i$ (deg) & $\Omega$ (deg) & $\omega$ \\ \hline 
        Starlink Phase 1 & 6921.8985 & 0.0000971  & 53.0532   & 227.0308        & 87.3346
    \end{tabular}
    \caption{Asset Satellite Orbit Parameters}
    \label{tab:starlinkORBParam}
\end{table}

%%%%%%%%%%%%%%%%%%%%%%%%%%%%%%%%%%%%%%%%%%%%%%%%%%%%%%%%%%%%%%%%%%%%%%%%%%%%%%%
\subsection{The KOSTPM Surfaces}
The representation of the family of transfer orbits parameterized by the velocity curve is convenient for studying the future space environment. The main idea is to obtain the surface of every possible velocity, $\Xi$, that creates a transfer orbit from the impact location to any true anomaly of the asset's orbit. Therefore, instead of having a single velocity hyperbola corresponding to a specific true anomaly, $\mbf v_0(\theta)$, the software will delineate a two-dimensional surface, $\Xi$, that covers the whole orbit: 
\begin{equation}
    \Xi = \cup_i\mbf v_0(\theta_i)\quad  \forall \quad 0<\theta<2\pi
\end{equation}
where  $\cup$ is the union operator. That is, the $\Xi$ surface is made by considering every velocity hyperbola for the full orbit revolution of the asset spacecraft. The threat assessment of the break-up event for a specific satellite comes by studying the intersection of the $\Xi$ surface with the region of velocities delimited by the NASA Break-up model, identified with the hulls. If an overlapping exists, that means that some of the debris generated by the collision or explosion will cross the satellite's pathway, and they represent a potential hazard that the asset needs to be aware of and plan collision avoidance maneuvers. 

\begin{figure}[!htb]%
\centering
\subfigure[View 1.]{%
\label{ko1}%
\includegraphics[width=.5\linewidth]{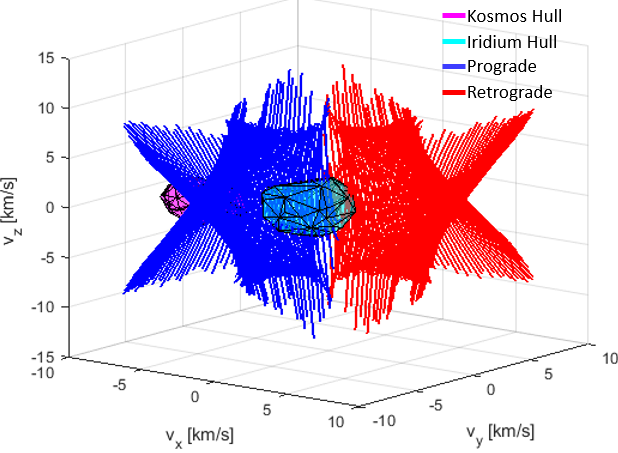}}%
\subfigure[View 2.]{%
\label{ko2}%
\includegraphics[width=.5\linewidth]{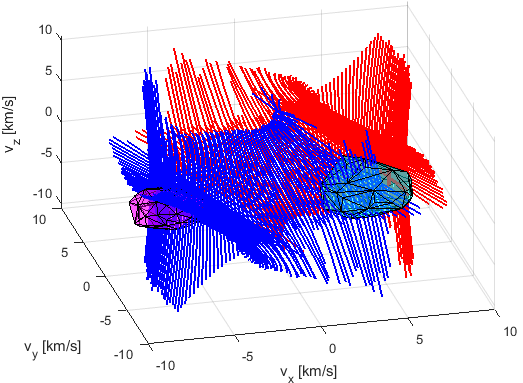}}%
\caption{KO threat evaluation. Transfer orbit surfaces and debris hulls.}
\label{KOcones}  
\end{figure}

Figure \ref{KOcones} reports the $\Xi$ surface for the proposed application, as well as the two hulls of debris related to the Iridium and Cosmos families. The surface $\Xi$ is made of two separate cones, which correspond to the side on which the debris is approaching the asset's orbit, either from below or above. However, the figure shows a vertical division of the cones, marked in different colors, red and blue. The blue velocity curves are connected to prograde transfer orbits, while red hyperbolas represent retrograde orbits. The KOSTPM solves for the full set of functions, but it needs to divide them into the two sets according to their motion. Surface $\Xi$ in the figure is created by plotting multiple hyperbolas, one next to the other. In this proposed application, a hyperbola every two degrees has been reported. It can be noted that the accuracy of the KO solution degrades when close to the switching section from prograde to retrograde solutions, as the analytical approach has difficulties in highlighting one behavior over the other. 

The hulls of debris are reported in the figure as well. The NASA SBM has been simulated for $10^4$ times, generating more than $10^8$ total debris. Therefore, by covering every possible outcome from the collision event, the hulls of debris represent a worst-case scenario application. The crossing of $\Xi$ inside the hulls informs that some debris will change their velocity and set themselves in an orbit that represents a threat to the asset satellite, as the pathways will cross. Consequently, the asset needs to be informed of the occurrence of the collision since the new orbits of debris will keep orbiting in a shared region of space and represent a continuous threat. 

Contrary to normal transfer maneuvers, where a spacecraft changes its orbit or docks after reaching the desired final position, the debris from the collision event will continue orbiting in the new orbits, posing a long-term threat to the satellite. Being in the LEO environment, the KO solution considers the nodal precession of the debris' orbit, and therefore the intersection region in the velocity domain previously described shifts as time proceeds. Thus, instead of deriving the $\Xi$ cones from the solution of Eq. \eqref{eq_kozeros}, the KO analytical approach can account for multiple revolutions by deriving the velocity hyperbolas from Eq. \eqref{eq:multi}. The surface $\Xi$ would shift in the velocity domain according to the $J_2$ nodal precession.

%%%%%%%%%%%%%%%%%%%%%%%%%%%%%%%%%%%%%%%%%%%%%%%%%%%%%%%%%%%%%%%%%%%%%%%%%%%%%%%
\subsection{The DASTPM Surfaces}
\begin{figure}[ht]
    \centering
    \includegraphics[width=0.8\textwidth]{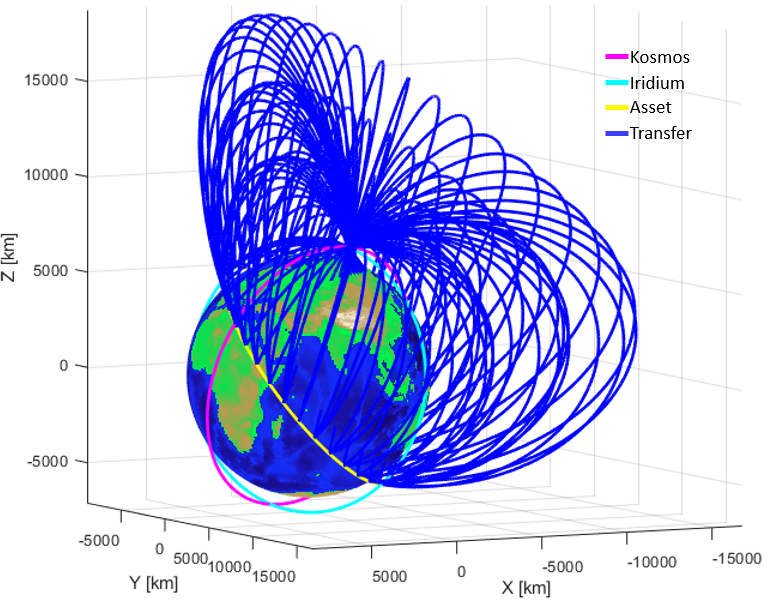}
    \caption{DA Solution, Side A of transfer orbits, $10^\circ$ separation.}
    \label{DAside}
\end{figure}

The DA solution of the $\Xi$ surface follows a different approach than the KO counterpart. Rather than solving for the prograde and retrograde velocity hyperbolas, the DASTPM, working with deviations, solves the upper side and the lower side transfer orbits independently, deriving the $\Xi$ cones separately. Figure \ref{DAside} shows the family of transfer orbits connected to Side A, where the debris reaches the asset from above. Indeed, Side A orbits are those that intersect the asset orbit coming from above. The figure shows a few orbits from each velocity hyperbola, with a different DASTPM evaluated every 10 degrees in true anomaly of the asset orbit, reported in yellow. Side B of the solution is not reported in the figure, but it would cover the second cone of $\Xi$, the one related to velocity hyperbolas connected to transfer orbits that cross the asset's orbit with the debris reaching from below. Figure \ref{DAside} shows the different approache used by DA to solve the same problem. Working with deviation from previously derived solutions, it is easier to compute all the hyperbolas on one side of the asset orbit since the curves are close to each other in the velocity domain. 

\begin{figure}[ht]
    \centering
    \includegraphics[width=0.8\textwidth]{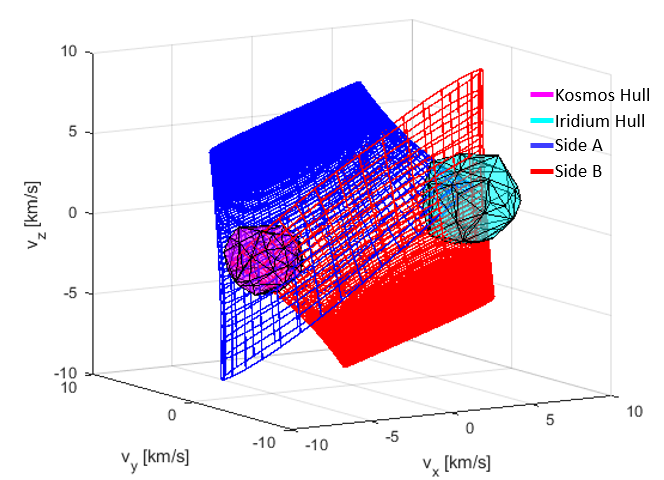}
    \caption{DA threat evaluation. Transfer orbit surfaces and debris hulls.}
    \label{DAcones}
\end{figure}

Similarly to the KO visualization, Fig. \ref{DAcones} reports the two cones of possible velocities that intersect the break-up point with the asset orbit. With the DA solution, each cone is one single color, one for Side A and another for Side B, as the DASTPM solves the velocity hyperbola following a different approach than the KO solution, where there is no need to know if the orbit is prograde or retrograde. Similarly to the KO figure, the hulls intersect the cone, meaning that the Iridium-Kosmos collision is a threat to the asset satellite. 

Figure \ref{DAcones} and Fig \ref{KOcones} report the same surface evaluated following two separate methodologies. However, the cones are the same and they would overlap if plotted together in the same graph. While the DA solution looks more clean and organized, it is worth noticing that the KO solution better describes the tails of the cones. Indeed, the DASTMP is set to work given a specific time of flight. On the contrary, the KO curves cover the full velocity domain from boundary to boundary due to the analytical nature of the KO derivation of the velocity hyperbolas. As previously mentioned, this representation of the debris covers most worst-case scenarios due to the large volumes of the hull compared to the actual one. This aspect is due to the elevated number of Monte Carlo simulations run to create every possible debris velocity offset.

%%%%%%%%%%%%%%%%%%%%%%%%%%%%%%%%%%%%%%%%%%%%%%%%%%%%%%%%%%%%%%%%%%%%%%%%%%%%%%%
%%%%%%%%%%%%%%%%%%%%%%%%%%%%%%%%%%%%%%%%%%%%%%%%%%%%%%%%%%%%%%%%%%%%%%%%%%%%%%%
%%%%%%%%%%%%%%%%%%%%%%%%%%%%%%%%%%%%%%%%%%%%%%%%%%%%%%%%%%%%%%%%%%%%%%%%%%%%%%%
\section{Conclusions}
Combining the distribution of debris generated from a break-up event in LEO with the transfer problem of identifying every possible orbit connecting two points in space leads to a fast visual representation of the possible threat posed by a close conjunction to an asset spacecraft orbiting independently. This problem has been approached from two separate directions that lead to the same results. The Koopman Operator methodology showed its strength in providing an analytical solution to the velocity hyperbola where the analytical curve is identified by funding the zeros of a function. The KO technique utilizes the true anomaly as the independent variable, which is constant given two position vectors in space. Thus, there is no need to specify different times of flight since the KO solution covers every possible transfer orbit. On the other hand, the DA approach makes use of the polynomial map inversion to obtain the velocity of the transfer orbit point-wise, efficiently solving multiple two-point boundary value problems. 

The QuickHull algorithm has been selected to represent the outcomes of the NASA Standard Break-up Model. Two convex hulls of velocities enclosing every possible orbit of the debris generated by the collision are identified. Following the SBM, two separate hulls are created since generated debris follow the motion of their parent RSO, before collision. By merging together the information derived from the KO and DA surfaces and the hulls, the threat evaluation is completed by assessing if there is any intersection between the surfaces. 

This paper focuses on the mere evaluation of threat providing a binary dangerous or not-dangerous outcome. Future work includes providing the degree of the threat in the case of a positive outcome, where the area of intersection between the surfaces is evaluated and a level of probability is given. Moreover, the hull not uniform probability of debris can also be taken into account to obtain a more accurate likelihood of pathway intersection. Lastly, some of the orbits reported in $\Xi$ might not be physical, especially the ones connected to the ``below" cone. The proposed techniques identify every possible orbit without considering the dimensions of the Earth. Future development will remove the parts of the cones of orbits that reach the asset's orbit going through the Earth's surface, meaning that those debris are actually deorbiting and represent no actual threat.

\bibliography{references.bib}

\end{document}